# Inline AI: Open-source Deep Learning Inference for Cardiac MR

Hui Xue[1], Rhodri H Davies[2], James Howard[3], Hunain Shiwani[2], Azaan Rehman[1],
Iain Pierce[2], Henry Procter[4], Marianna Fontana[5], James C Moon[2],
Eylem Levelt[4], Peter Kellman[1,2]

1. National Heart, Lung and Blood Institute, National Institutes of Health, Bethesda, MD, USA
2. Barts Heart Centre, Barts Health NHS Trust, London, UK
3. Hammersmith Hospital, London, UK
4. University of Leeds Hospital, Leeds, UK
5. National Amyloidosis Centre, Royal Free Hospital, London, UK

# Corresponding author:

Hui Xue

National Heart, Lung and Blood Institute
National Institutes of Health
10 Center Drive, Bethesda
MD 20892
USA

Phone:  +1 (301) 827-0156
Cell:   +1 (609) 712-3398
Fax:    +1 (301) 496-2389
Email:  hui.xue@nih.gov

**Word Count: 3,030**

hui.xue@nih.gov
rhodri.davies2@nhs.net
james.howard@nhs.net
hunain.shiwani@nhs.net
azaan.rehman@nih.gov
iain.pierce@nhs.net
h.procter@leeds.ac.uk
m.fontana@ucl.ac.uk
j.moon@ucl.ac.uk
e.levelt@leeds.ac.uk
kellmanp@nhlbi.nih.gov

# Inline AI: Open-source Deep Learning Inference for Cardiac MR

## Key points:

1. Proposed a new imaging AI framework, InlineAI, to allow model deployment into imaging workflow in a streaming manner directly on the scanner. Users can extend the system by plugging in their models.

2. InlineAI consists of configurable image reconstruction and image analysis with flexible deep learning model inference support, where key biomarkers are measured on incoming images with full automated and available for evaluation while imaging is ongoing.

3. InlineAI capacity was demonstrated on three cardiac MR applications for applicability, inference speed and supported output format.

## Summary statement:

An Inline AI open-source framework is developed to enable the effective model inference in the imaging workflow on the scanner, allowing biomarker extraction and reporting with full automation.

**Abbreviations**

CMR = cardiac magnetic resonance
DL = deep learning
EF = ejection fraction
LV = left ventricular
ESV = end-systolic volume
EDV = end-diastolic volume
SV = stroke volume
CO = cardiac output
GPU = graphical processing units
CH4 = four-chamber (CH4)
CH2 = two-chamber (CH2)
GL-Shortening = global longitudinal shortening
MAPSE = mitral annular plane systolic excursion
TAPSE = tricuspid annular plane systolic excursion
AIF = arterial input function

**Abstract**

**Background**

Cardiac Magnetic Resonance (CMR) is established as a non-invasive imaging technique for evaluation of heart function, anatomy, and myocardial tissue characterization. Quantitative biomarkers are central for diagnosis and management of heart disease. Deep learning (DL) is playing an ever more important role in extracting these quantitative measures from CMR images. While many researchers have reported promising results in training and evaluating models, model deployment into the imaging workflow is less explored.

**Methods**

A new imaging AI framework, the InlineAI, was developed and open-sourced. The main innovation is to enable the model inference inline as a part of imaging computation, instead of as an offline post-processing step and to allow users to plug in their models. We demonstrate the system capability on three applications: long-axis CMR cine landmark detection, short-axis CMR cine analysis of function and anatomy, and quantitative perfusion mapping.

**Results**

The InlineAI allowed models to be deployed into imaging workflow in a streaming manner directly on the scanner. The model was loaded and inference on incoming images were performed while the data acquisition was ongoing, and results were sent back to scanner. Several biomarkers were extracted from model outputs in the demonstrated applications and reported as curves and tabular values. All processes are full automated. the model inference was completed within 6-45s after the end of imaging data acquisition.

**Conclusions**

An Inline AI open-source framework is developed to enable the effective model inference in the imaging workflow on the scanner. Its capability is demonstrated on three CMR imaging applications.

**Key words**

# Introduction

Cardiac Magnetic Resonance (CMR) imaging is established as a mainstream non-invasive imaging technique for evaluation of heart function, anatomy and myocardial tissue characterization. It is the gold-standard for measurement of cardiac function used in clinical decision-making (1). In the past decade the measurement of CMR biomarkers has relied on manual and semi-automated image analysis by the expert cardiologists (2), which is time-consuming and inconsistent, reducing precision (3).

Deep learning (DL) is rapidly changing this landscape to enable automated biomarker extraction from CMR scans (4). A rich body of publications are in the literature to segment the cardiac chambers to measure functional biomarkers like ejection fraction (EF) and left ventricular (LV) mass (5). A recent study showed the automated DL models can surpass expert-level performance in measuring LV structure and global systolic function, using scan-rescan experiments (6). As well as cardiac function, deep learning models have been further developed to analyze quantitative perfusion imaging (7), T1 mapping (8), LGE (9) and to automate strain analysis from tagging CMR (10). The overall trend is clearly towards fully automated analysis for CMR scans. DL models will play a central role here and the research along this direction remain very active (4).

While many researchers have reported very promising results in training and evaluating models in CMR, there are limited reports of such systems being used in the imaging workflow and later in the clinical care(11). Commercial software nowadays provides vendor-supplied deep learning models for different image analysis tasks (12), but these post-processing steps can be time consuming and often require specialized computing equipment (e.g. graphical processing units (GPU) ) on reporting workstations. Furthermore, there is limited support for integration of user-developed models. We propose an open-source system which allows users to deploy easily their DL models, to enable fast prototyping, refinement and distribution of models. For this purpose, this paper proposes and demonstrates an effective deep learning model inference system, called Inline AI, for cardiac MR imaging which can be performed directly on the scanner.

After a model is trained on the retrospectively curated datasets, it is deployed into the imaging workflow and applied to new images for inference (Figure 1). The inference results, such as segmentation contours or landmarks, are displayed directly on the scanner console and reviewed while the scan is ongoing. These results, together with measured biomarkers are saved in the DICOM images. The entire model inference process is automated, without disrupting the imaging workflow.

A system allowing the effective model inference has benefits. First, it will allow model to be continuously evaluated on new data, which serves as a real-world testing for the model performance. Model performance drifting caused by e.g. different pathological condition, patient cohort or imaging variation, can be exposed and may potentially be fixed by further training. Second, the automated nature of DL models on CMR analysis enables auto-reporting for key biomarkers (as illustrated later in the paper). The Inline AI setup will streamline the image analysis and biomarker reporting and has potential to reduce reporting time. Third, the fast feedback loop of model inference and biomarker reports can serve as the quality assurance facilitating repeated acquisitions if required. For example, if model segmentation is failed due to a mis-placed imaging plane, this error can be spotted by the imaging technician by checking the model outputs and may get corrected in time. Fourth, since the model inference is already performed inline, the clinicians have more freedom to conduct reporting on less-powerful computers or mobile devices, without powerful CPU or GPU.

The InlineAI system is built as an extension to the open-source framework Gadgetron (13). It consists of configurable image reconstruction and image analysis with DL model inference support. Users can extend the system by plugging in their models. We demonstrate the system capability on three applications: long-axis CMR cine landmark detection, short-axis CMR cine analysis of function and anatomy, and quantitative perfusion mapping.

## Methods

**Architecture and system design**

InlineAI is built on top of the open-source Gadgetron framework (14) (Figure 2). The Gadgetron reconstruction process consists of three components: Readers, Writers, and Gadgets. A Reader receives and deserializes the incoming data sent from the client (e.g., a client can be the MR scanner). Examples of data can be MR Kspace readouts, images or text. A Writer serializes the reconstruction results and sends the data packages to the client. The Gadgets are the programmable components to take in the input data and apply processing. The Gadgetron, running as a server, communicates with clients. The client will initialize the connection to the server by sending a configuration file, stating which readers, writers and gadgets are required in a specific order. Gadgetron will dynamically load and assemble the processing chain (aka, gadget chain). While client streams the data, Gadgetron will buffer and process data on the fly.

The processing results after model inference are sent back to client. After the processing is completed, Gadgetron will shut down the connection and release any resources (such as CPU, GPU, RAM etc.).

To facilitate the flexible DL imaging applications, a set of dedicated Gadgets were developed and added to Gadgetron, as shown in Figure 3. Following streaming image reconstruction, images are organized and analyzed, including loading and applying the DL models.

The reconstruction stage features a few key steps. Accumulation and triggering: This step includes *KspaceBuffering* and *Triggering* gadgets. The main function is to buffer the incoming kspace data and send out the buffered data to next stage if a triggering condition is met. Here the triggering condition determines when to send out the buffered data. For example, a multi-slice cine imaging can be triggered by slice index. It means if the kspace data from the 2nd slice arrives, it indicates all data from the 1st slice had been acquired and buffered data can be sent to next gadget (by setting the *trigger_dimension* to *slice* in the *Triggering* gadget). This feature is important to allow data acquisition and computing to be performed in parallel; otherwise, computing has to wait until the entire scan is completed, which will cause unnecessary delay. Reconstruction: This step includes *PrepareRefData* and *Reconstruction* gadgets. The *PrepareRefData* assembles the reference calibration data for parallel imaging, if needed. The *Reconstruction* gadget converts the buffered kspace data to images. If the dynamic triggering per slice was used, reconstruction will be performed once per slice, while the data acquisition is ongoing for new slices. Gadgetron has implemented a number of reconstruction algorithms, such as GRAPPA (15) and L1-SPIRIT (16) etc. These algorithms are interchangeable by inserting different gadgets at this step.

The image analysis stage includes gadgets to organize the images and conduct analysis. The *ImageBuffering* gadget will first buffer incoming images, where user can specify the condition to start analysis. For example, one application may require all images (e.g., cardiac phases) for a slice to be processed together. *ImageAnalysis* gadget will receive buffered images and perform model inference. As shown in Figure 3, the framework provides the python-C++ interface based on Boost.Python (17), since Python is the language of choice used by major deep learning frameworks (18). The framework allows user to supply python functions for loading and applying AI models. Often, models are loaded once in the configuration stage of gadgets and applied to incoming data repeatedly. The implementation currently supports both Pytorch (19) and ONNX (https://onnx.ai/) format of models. User can specify whether inference happens in CPU or GPU, depending on the runtime environment. Python-C++ data conversion is implemented for all major MR data types, including k-space, image, ECG/respiratory waveform, XML meta data and contours and anatomical landmarks.

A complete example including model loading and inference can be found in Gadgetron repository (https://github.com/gadgetron/gadgetron/blob/master/gadgets/cmr/config/LandmarkDetection/gadgetron_cmr_landmark_detection.py).

**InlineAI applications**

To demonstrate the concept and capacity of Inline AI, model inference was implemented for three example applications. Models for each application had been previously published and validated in prior studies; this paper focuses on inline integration. Using InlineAI, we were able to deploy these models into the imaging workflow and achieve full automation in analysis and biomarker reporting.

Longitudinal function analysis from cine images

This application aims to detect the mitral and tricuspid valve points and apical points from the long axis cine images. The detected landmarks are used to compute global longitudinal shortening (GL-Shortening), mitral annular plane systolic excursion (MAPSE), and tricuspid annular plane systolic excursion (TAPSE). The estimated biomarkers were tested on an independent cohort for prognostic prediction power and found to have strong associations with adverse outcomes. Details of this study can be found in previous publications (20,21).

Models were integrated inline using the proposed framework. The pre-trained models were saved to ONNX format for CPU inference and converted to Pytorch Script format for GPU inference. The reconstructed cine image series were input into the model inference and landmarks were detected on every phase and plotted on the images (Figure 4). GL-Shortening and MAPSE were computed and reported as plots and tables.

Short axis cine analysis

This application aims to analyze short-axis cine images and measure cardiac size, mass, and function. All image acquisition reflected standard international recommendations (2). The input is the stack of SAX cine images and models are used segment the endocardial and epicardial boundary for end-diastolic and end-systolic phases. Based on the segmentation, the EF, LV MASS, end-systolic volume (ESV), end-diastolic volume (EDV), stroke volume (SV) and cardiac output (CO) were computed. The myocardial contours were derived from the segmentation outputs. The model showed better precision on the scan-rescan dataset compared to an expert cardiologist. Further details of this study can be found in a previous publication (6).

For inline implementation, models were trained using Pytorch and converted to ONNX format for CPU based inference. After the image reconstruction, retro-gated cine images were input to image analysis. To parallelize the multi-slice imaging with computing, the reconstruction was triggered once for every slice. The models were applied to segment LV blood-pool and myocardium. The boundary contours were computed from the segmentation mask. The ED and ES phases were identified with the largest and smallest blood volume and used to compute biomarkers. The inline analysis further created mosaiced panels of ED and ES images (Figure 5), with overlaid segmentation contours for quick review. The valve plane at both ED and ES phases is determined from the inline analysis of long axis cine images as described above. The valve plane intersection with the SAX images is used to determine the extent of the LV, and therefore which slices are subsequently used in computation of mass and volumes. Previously computed long axis data is temporarily stored within Gadgetron and available for the SAX analysis. The computed biomarkers are presented as report pages, saved into the DICOM database.

Quantitative Myocardial Perfusion mapping analysis

Quantitative perfusion mapping is an emerging technique to produce pixel-wise myocardial flow maps at rest or stress. Compared with qualitative visual assessment, quantitative mapping is more objective and sensitive. Recent progress enables automated mapping and generation of pixel-wise perfusion flow maps on the scanner (23,24). This opens the door for an end-to-end analysis from the images to myocardial perfusion biomarkers, using deep learning. A deep CNN model was thus developed for this purpose (7). The myocardial stress flow measured by model was found to be a strong, independent predictor of adverse cardiovascular outcomes (25). The inline analysis further extracts the arterial input function (AIF) signal from the RV and LV blood pool, from which the pulmonary transit time is computed (26).

Perfusion models were trained in Pytorch and exported in ONNX format for inference. The image analysis gadget first loads the model and applies it to the 2D+T perfusion images. The RV blood pool segmentation was used to determine the RV insertion point and myocardial segmentation was split into sectors using the AHA 16 sector model. The sector-wise myocardial blood flow is displayed as a bullseye plot and reported in a table (Figure 6). The sector segmentation and myocardial blood flow were further split into endo and epi layers. To relate the rest and stress perfusion scans, the analysis chain was extended to link two scans. From these, the myocardial perfusion reserve report was computed. The scan information, such as patient heart rate and respiratory condition, are included in the report. The complete process is fully automated, serving as a one-click solution for inline perfusion analysis.

**Imaging experiments**

All imaging experiments were conducted with National Research Ethics Committee (Ref:18/YH/0168) approval. The study complied with the Declaration of Helsinki and written informed consent was obtained from each participant.

Inline integration tests were performed at the XXX hospital on a clinical scanner (3.0T MAGNETOM Prisma, Siemens AG Healthcare, Erlangen, Germany). Patients underwent a standard cardiac MR exam acquiring both SAX and LAX cines images using Inline AI. For the SAX scan, a total of 11 parallel slices were prescribed to cover the LV. Breath-holding was ~7s per slice, with a 4s gap, resulting in the total imaging time of ~2.4mins. For the LAX analysis, CH4 and CH2 slices were prescribed and imaged. To demonstrate the perfusion application, an adenosine stress test with the inline perfusion analysis application was used for evaluating myocardial stress perfusion.

All imaging experiments were conducted with local institutional approval and written informed consent. A timing test was performed inline on an external networked Linux server (AMD EPYC 7763 64-Core Processor, 64GB RAM, NVIDIA GeForce RTX 3090 with12 GB GPU RAM).

## Results

Figure 4 illustrates the landmark detection for inline calculation of LAX function. For the CH4 view, the mitral valve point and apical point were detected for every phase. The myocardial length, and the global longitudinal shortening were plotted. Measured biomarkers are reported in a table. The tricuspid valve points were further detected from the CH4 view and resulting

TAPSE was computed. For the CH2 view, the mitral valve and apical points were detected for every phase. The computing time was 12s for the CPU inference and 6s if using the GPU.

Figure 5 illustrates the Inline SAX analysis results that are output in addition to the actual cine images. Data acquisition was overlapped with computing, with reconstruction and analysis being triggered by every slice. That is, following the streaming image reconstruction, the model was processing the first slice while the second one was still being imaged. After all imaging slices were acquired, the model segmentation was accumulated and analyzed. For the ED phase, the endocardial and epicardial boundaries were detected and rendered in a mosaic panel, for rapid evaluation. The per-slice contribution is listed in the mosaic image panel for LV volume and mass. For the ES phase, the endocardial boundaries were detected and rendered. By combining the ED and ES segmentation, the inline analysis computes and generates the report page, including EF, EDV, ESV, SV, MASS and cardiac output (CO). The corresponding index values are given by normalizing with the body-surface-area (BSA). Furthermore, the per-slice maximal wall thickness at ED phase was measured and presented in a third mosaic panel. The computing time measured from the end of data acquisition to the moment all results were sent back to the scanner was 35s, with full automation, which includes the retrogated image reconstruction, inline analysis, and generation of all output series. Without the on-the-fly reconstruction triggered by every slice, the computing will not start until all slices were acquired lengthening by an extra 1.5mins.

Figure 6 illustrates the quantitative perfusion inline analysis application. Pixel-wise myocardial perfusion flow maps are generated for all three SAX slices. The model segmented the perfusion scan and myocardium was split according to the standard AHA 16-segment model, with the RV insertion correctly identified. The per-sector contours were rendered on the perfusion maps for visual evaluation. From the per-sector segmentation, the mean myocardial flow was reported in the AHA bulls-eye plot and in tabular format. The report was further refined to give per-sector endocardial and epicardial flow values. The myocardial perfusion reserve results were presented as another bulls-eye report, dividing the stress flow by the rest for each sector. The segmentation, AHA report and other results are displayed immediately on the scanner. The model inference was performed on the CPU with a computing time of only 250ms per slice.

## Discussion

This paper presents an InlineAI open-source extension to the Gadgetron framework to allow effective model inference as a part of imaging computation. The key innovation is to enable model deployment into the imaging workflow and allow models to be applied on-the-fly with efficiency and flexibility. Users can plug in their models into the imaging workflow. The inference results are displayed on the scanner console for reviewing, with full automation, while the patient scan is on-going. As demonstrated by three examples, different outputs are supported, including contours, landmarks, and biomarkers. This allows the scanner operators and later reporting clinicians to effectively evaluate AI performance. The framework allows integrating pre-developed deep learning source code (e.g. in Python) with minimal revision. The reconstruction allows flexible triggering to parallelize imaging and computing.

InlineAI integration was demonstrated on three applications: SAX cine analysis for cardiac function, LAX landmark detection-based GL-Shortening and MAPSE measurement and quantitative perfusion. For each example, the models were pre-developed on sizable offline curated datasets and validated extensively including outcome studies of prognostic value (6,21,25,26). With InlineAI, we demonstrated these pre-trained models are integrated into the CMR imaging workflow. The reports of measured biomarkers are immediately available. In addition to the image displays, the contour and report data is stored as metadata in the DICOM headers for off-line retrieval and retrospective analysis.

Inline model inference has advantages. Faster feedback is achieved by streamlining the model inference, as the model outputs are available during the imaging session for evaluation. The scanner operator has the chances to rescan or add more imaging, based on the inline outputs of previous scans. The availabilities of model biomarkers can encourage data driven imaging. One example is the globally elevated native T1 values reported by a model may indicate fibrosis and trigger a post-contrast mapping for ECV measurement. On the reporting side, the inline biomarkers are readily available to reduce the measurement workload of clinicians. The auto-measured biomarkers can be useful to optimize reporting priorities by providing clues for disease severity.

## Conclusions

This paper presented an Inline AI open-source framework to enable the effective model inference in the imaging workflow on the scanner and demonstrated its capability on three previously reported CMR imaging applications.

**Declarations**

**Ethical Approval and Consent to participate**

All imaging experiments were conducted with National Research Ethics Committee (Ref:18/YH/0168) approval. The study complied with the Declaration of Helsinki and written informed consent was obtained from each participant.

**Consent for publication**

Data was acquired with the required ethical approval. The IRB approval document is available for review by the Editor-in-Chief.

**Availability of data and material**

The data supporting the findings from these investigations are available within the article and the supplementary material or are available upon reasonable request to the authors. The software is openly shared at the github repo: https://github.com/gadgetron/gadgetron

**Competing interests**

The authors declare that they have no competing interests

**Funding**

-

**Authors' contributions**

-

# List of Captions

**Figure 1** System schematic plot. The InlineAI is built upon the open-source Gadgetron framework. It receives streamed data from MR scanners and performs image reconstruction and applies models to images for inference. The model outputs and measured biomarkers are sent back to the scanner and stored in the typical DICOM database, with full automation. The reporting clinicians are in the loop to examine the model performance and decide to accept the measurements. This high level of automation speeds up the reporting process.

**Figure 2** Illustration of the Gadgetron workflow. The running service consists of readers, writers and a set of programmable components, called Gadgets. A reader receives data from the scanner and a write sends back images and other results back. The Gadget chain is configurable with one or more gadgets calling deep learning models.

**Figure 3** An instantiation of Gadgetron chain to support model inference on images. Users are required to supply minimal python program to load and apply models. The framework supports dynamic triggering for reconstruction and model inference.

**Figure 4** Long-axis cine Landmark detection. The scanner screen snapshots for a CH4 cine scan are shown here. (a) The CH4 cine with detected landmarks overlaid on the images are shown. (b) The measured biomarkers from the CH4 cine is reported as a table. Since the landmarks were detected on all phases, the LV length (c) and the GL-Shortening (d) are presented as time curves against the trigger time.

**Figure 5** Short-axis cine analysis with the InlineAI. The scanner screen snapshots are shown here. (a) The ED panel includes all imaged slices with both endocardial and epicardial segmentation contours overlaid on images. (b) The ES panel is plotted with endocardial contours, which allows to measure the key functional biomarkers. (c) The measure biomarkers are reported as a table and sent back to the scanner. (d) The inline analysis further measures the maximal wall thickness for every slice. (e) The cross-reference intersections of SAX slices were plotted on LAX ED and ES phases, together with detected landmark points.

**Figure 6** Quantitative perfusion mapping analysis. (a) The stress (the first row) and rest (the second row) perfusion maps are overlaid with per-sector segmentation. The AHA 16 sector bulls-eye plots are generated on the scanner for the stress (the left on the third row) and the rest (the middle on the third row) scans. The right panel on the third row gives the myocardial perfusion reserve, by dividing the stress with the rest flow values. (b) A report page is further generated to record the scan information, such as patient heart rate and respiratory condition etc. (c) The analysis extracts AIF signals plotted as curves for both RV and LV. The pulmonary transit time is computed from AIF curves. Stress AIFs are shown here.

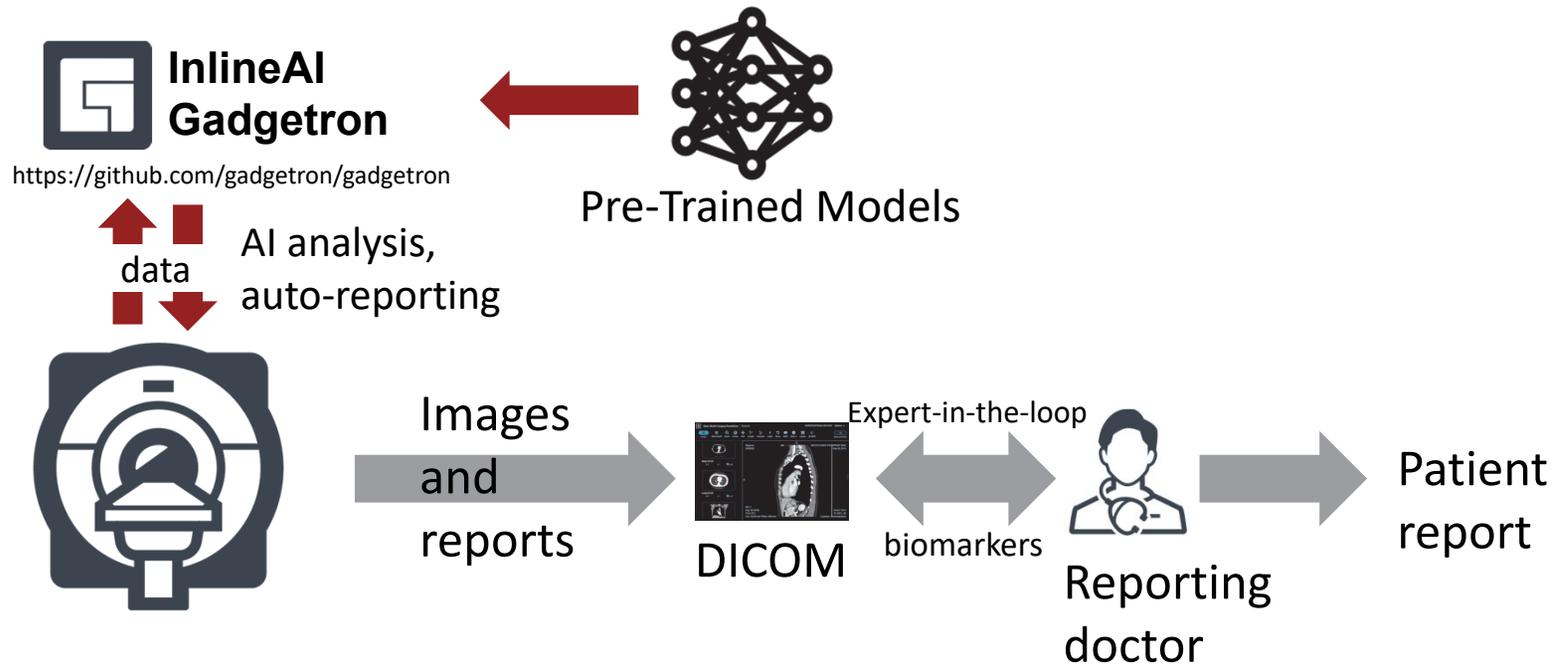

Figure 1. System schematic plot. The InlineAI is built upon the open-source Gadgetron framework. It receives streamed data from MR scanners and performs image reconstruction and applies models to images for inference. The model outputs and measured biomarkers are sent back to the scanner and stored in the typical DICOM database, with full automation. The reporting clinicians are in the loop to examine the model performance and decide to accept the measurements. This high level of automation speeds up the reporting process.

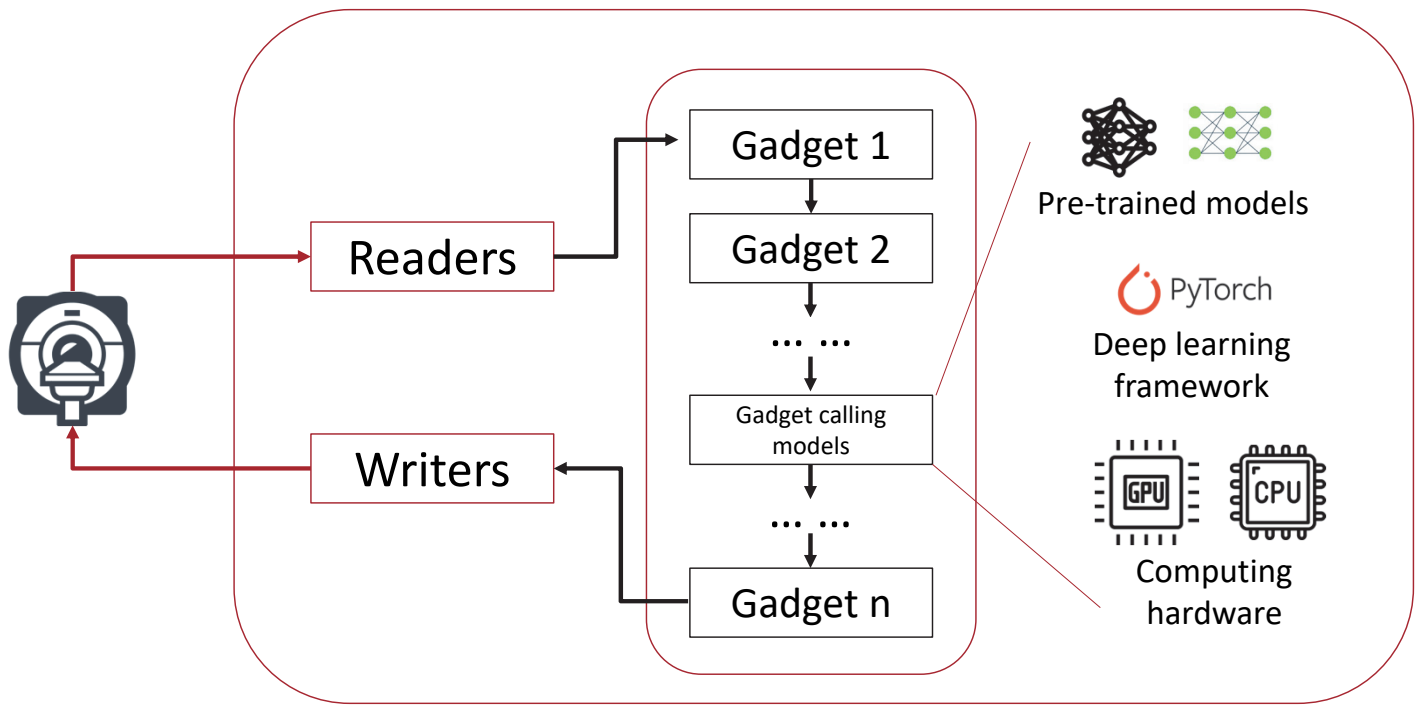

Figure 2. Illustration of the Gadgetron workflow. The running service consists of readers, writers and a set of programmable components, called Gadgets. A reader receives data from the scanner and a write sends back images and other results back. The Gadget chain is configurable with one or more gadgets calling deep learning models.

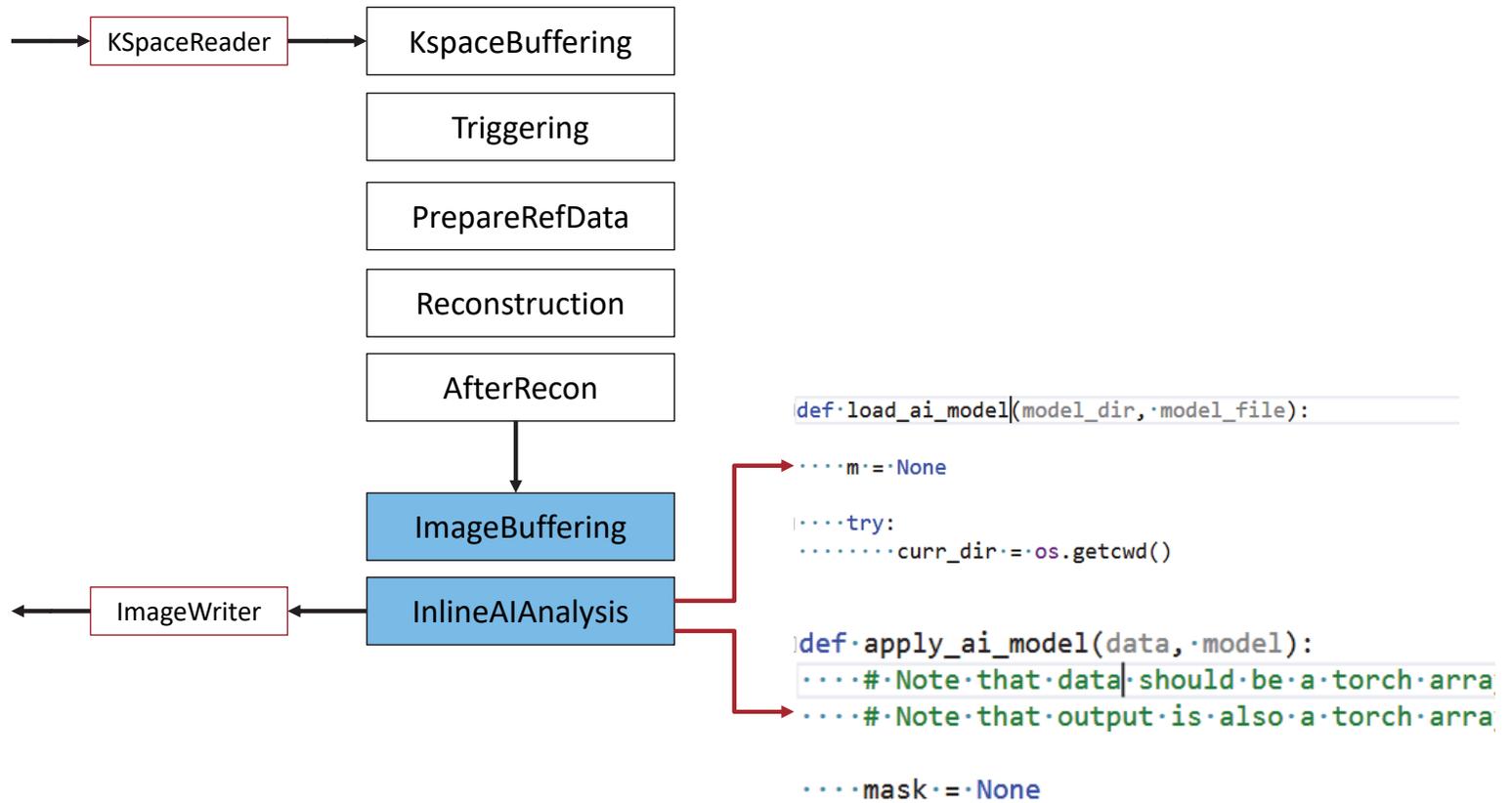

Figure 3. An instantiation of Gadgetron chain to support model inference on images. Users are required to supply minimal python program to load and apply models. The framework supports dynamic triggering for reconstruction and model inference.

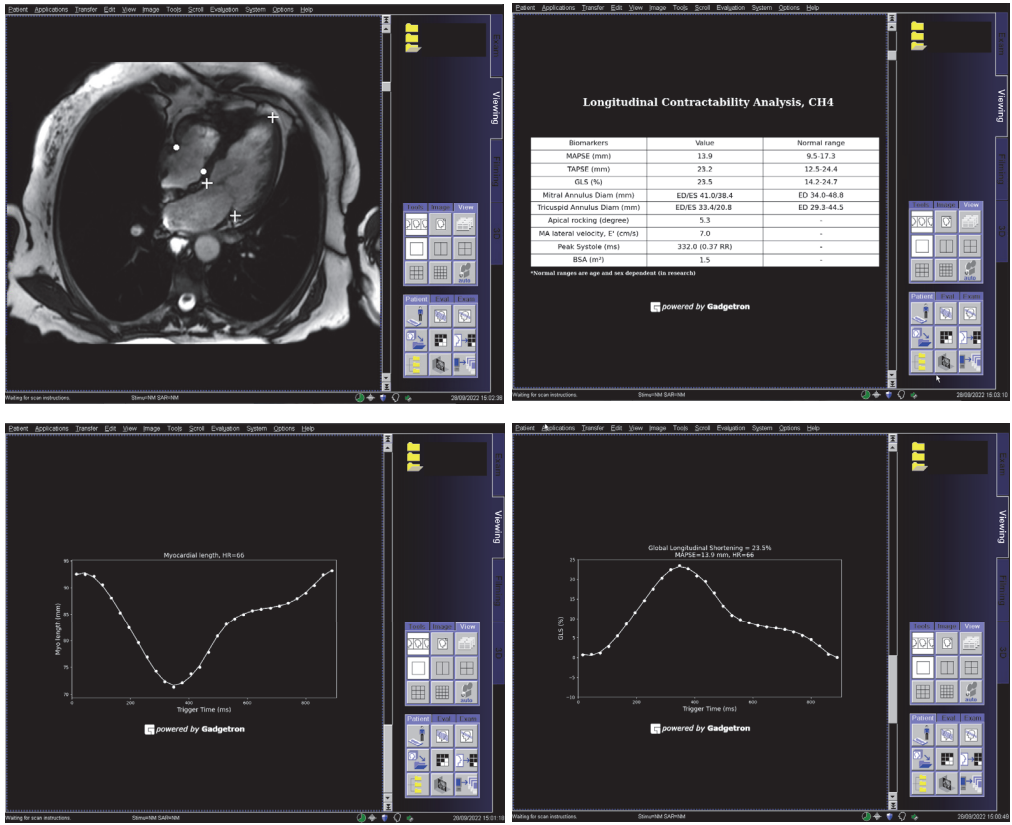

Figure 4. Long-axis cine Landmark detection. The scanner screen snapshots for a CH4 cine scan are shown here. (a) The CH4 cine with detected landmarks overlaid on the images are shown. (b) The measured biomarkers from the CH4 cine is reported as a table. Since the landmarks were detected on all phases, the LV length (c) and the GL-Shortening (d) are presented as time curves against the trigger time.

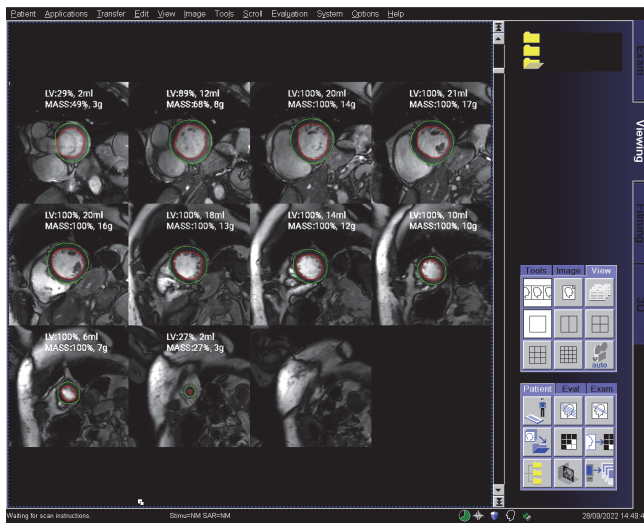
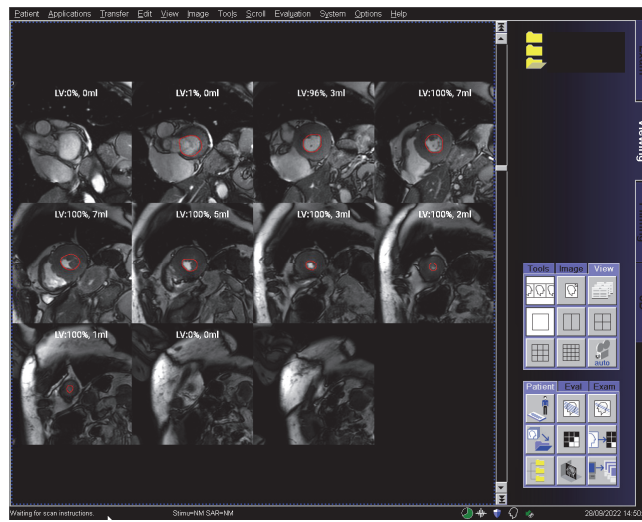
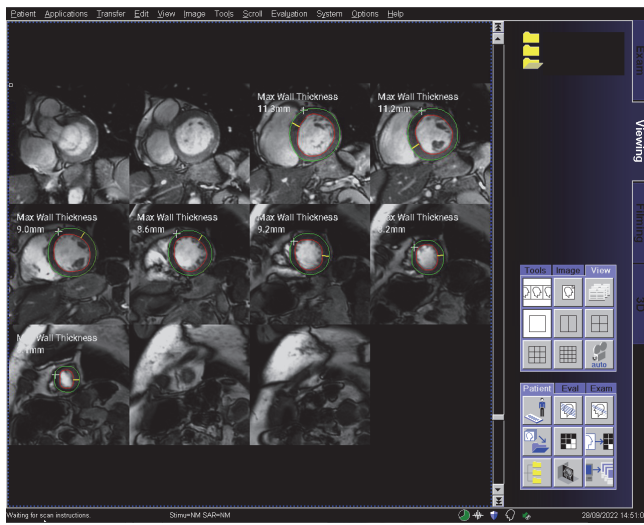
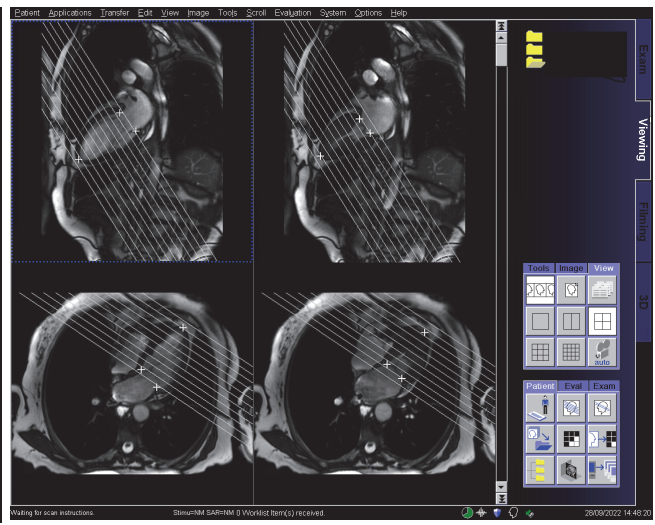

Figure 5. Short-axis cine analysis with the InlineAI. The scanner screen snapshots are shown here. (a) The ED panel includes all imaged slices with both endocardial and epicardial segmentation contours overlaid on images. (b) The ES panel is plotted with endocardial contours, which allows to measure the key functional biomarkers. (c) The measure biomarkers are reported as a table and sent back to the scanner. (d) The inline analysis further measures the maximal wall thickness for every slice. (e) The cross-reference intersections of SAX slices were plotted on LAX ED and ES phases, together with detected landmark points.

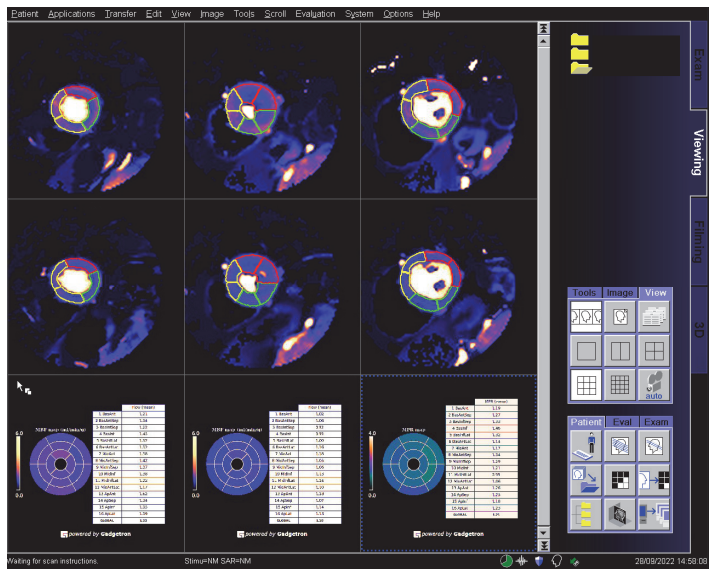
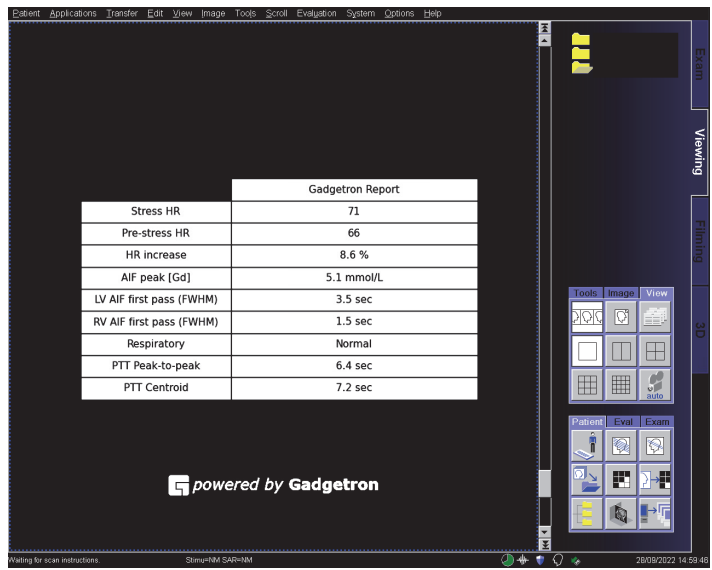
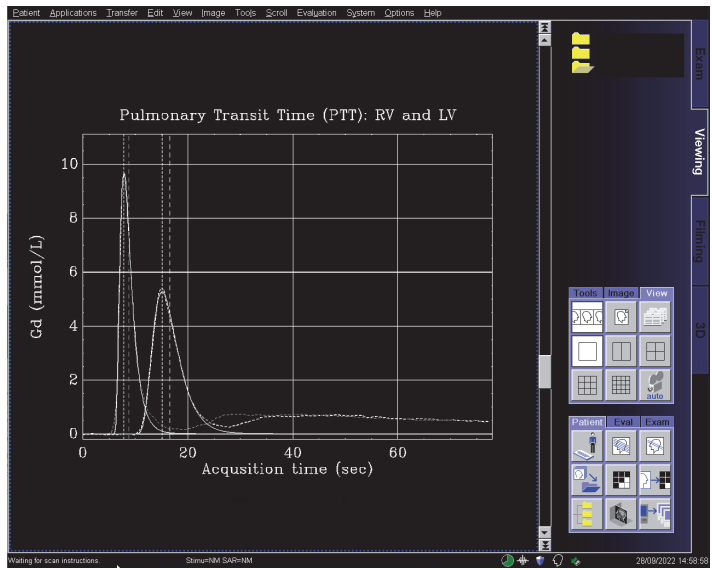

Figure 6. Quantitative perfusion mapping analysis. (a) The stress (the first row) and rest (the second row) perfusion maps are overlaid with per-sector segmentation. The AHA 16 sector bulls-eye plots are generated on the scanner for the stress (the left on the third row) and the rest (the middle on the third row) scans. The right panel on the third row gives the myocardial perfusion reserve, by dividing the stress with the rest flow values. (b) A report page is further generated to record the scan information, such as patient heart rate and respiratory condition etc. (c) The analysis extracts AIF signals plotted as curves for both RV and LV. The pulmonary transit time is computed from AIF curves. Stress AIFs are shown here.